\documentclass{article}

\usepackage{arxiv}

\usepackage[utf8]{inputenc} 
\usepackage[T1]{fontenc}    
\usepackage[hidelinks]{hyperref}       
\usepackage{url}            
\usepackage{booktabs}       
\usepackage{amsmath}        
\usepackage{amsfonts}       
\usepackage{nicefrac}       
\usepackage{cleveref}       
\usepackage{graphicx}
\usepackage[numbers]{natbib}
\usepackage{doi}
\usepackage[labelfont=bf]{caption}

\title{National and state-level datasets of United States forensic DNA databases 2001--2025}

\usepackage{authblk}

\newbox{\orcid}\sbox{\orcid}{\includegraphics[scale=0.06]{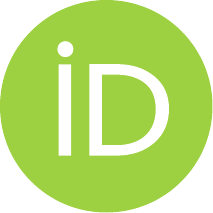}}

\author[1]{%
  \href{https://orcid.org/0000-0002-1942-786X}{\usebox{\orcid}\hspace{1mm}Yemko Pryor}%
}
\author[1]{%
  \href{https://orcid.org/0009-0005-5151-5068}{\usebox{\orcid}\hspace{1mm}Virum Ranka}%
}
\author[2]{%
  \href{https://orcid.org/0000-0002-6728-4373}{\usebox{\orcid}\hspace{1mm}Jo\~ao Pedro Donadio}%
}
\author[1]{%
  \href{https://orcid.org/0009-0007-8262-2872}{\usebox{\orcid}\hspace{1mm}Samantha C.~Muller}%
}
\author[3]{%
  \href{https://orcid.org/0009-0003-0667-4946}{\usebox{\orcid}\hspace{1mm}Jenna Wilson}%
}
\author[1,4,*]{%
  \mbox{\href{https://orcid.org/0000-0002-7546-6724}{\usebox{\orcid}\hspace{1mm}Tina Lasisi}}%
}

\affil[1]{Department of Anthropology, University of Michigan, Ann Arbor, USA}
\affil[2]{Independent researcher, Vancouver, Canada}
\affil[3]{Department of Human Genetics, University of Michigan, Ann Arbor, USA}
\affil[4]{Center for Ethics, Society, and Computing, University of Michigan, Ann Arbor, USA}
\affil[*]{Corresponding author: \texttt{tlasisi@umich.edu}}

\renewcommand{\shorttitle}{U.S. Forensic DNA Databases 2001--2025}

\hypersetup{
pdftitle={National and state-level datasets of United States forensic DNA databases 2001-2025},
pdfsubject={forensic genetics, CODIS, DNA databases, criminal justice},
pdfauthor={Pryor, Y., Ranka, V., Donadio, J. P., Muller, S.C., Wilson, J., Lasisi, T.},
pdfkeywords={forensic DNA databases, CODIS, NDIS, SDIS, United States, criminal justice},
}

\begin{document}
\maketitle

\begin{abstract}
Forensic DNA databases in the United States have expanded substantially over the past two decades. However, comprehensive, harmonized data describing database structure and composition remain limited. This dataset series documents forensic DNA infrastructure across national and state levels from 2001 to 2025. It includes a reconstructed time series of monthly National DNA Index System (NDIS) statistics from FBI archives, capturing counts of offender, arrestee, and forensic profiles, participating laboratory totals, and investigations aided. A complementary dataset compiles publicly available state-level statistics and policy metadata on arrestee collection laws, familial search practices, and DNA collection statutes across all 50 states. A third dataset provides standardized demographic and annual collection data obtained through previously published public records requests, including sex and racial composition where reported. Together, these resources provide a foundation for studying the historical development of forensic DNA systems in the U.S., enabling longitudinal and cross-sectional analyses of database growth, policy variation, and reporting practices across jurisdictions.
\end{abstract}


\section*{Background \& Summary}

The Combined DNA Index System (CODIS) is a software platform originated by the FBI to store, compare, and search DNA profiles using a standardized set of short tandem repeat (STR) loci \cite{Unknown2022-yi}. CODIS is implemented across three levels: local DNA index systems (LDIS) maintained by individual laboratories, state DNA index systems (SDIS) operated by state agencies, and the National DNA Index System (NDIS) administered by the FBI. DNA profile records flow upward from LDIS to SDIS to NDIS if they meet defined eligibility criteria and are organized into multiple indices, two of which are the Offender Index (which includes profiles from convicted individuals and, where authorized, arrestees) and the Forensic Index, which contains crime scene profiles of unknown offenders \cite{UnknownUnknown-zg}.

The FBI has published monthly NDIS statistics since the early 2000s, including counts of offender, arrestee, and forensic profiles, along with cumulative investigations aided \cite{Unknown2022-qe}. These statistics are commonly referenced in the literature, with studies reporting point-in-time totals based on the FBI's website at the time of writing. For example, counts from June 2011 \cite{Ge2012-if}, January 2012 \cite{Ge2012-ve}, May 2013 \cite{Ge2014-rl}, October 2021 \cite{Wickenheiser2022-tb}, November 2022 \cite{Link2023-dm}, and February 2024 \cite{Greenwald2024-sg}. However, these reports are released only as individual snapshots and are not preserved in a longitudinal, machine-readable format, limiting their use for tracking changes over time.

Data availability is inconsistent at the state level. Some states publish total SDIS counts or break down profiles by category, while others provide general department summaries or statutory references that are not specific to SDIS. Many do not publish SDIS-level statistics easily available to the public. Demographic composition is even more sparsely documented. The only known compilation of sex and racial composition data comes from a 2020 publication by Murphy and Tong \cite{Murphy2020-pc}, who submitted Freedom of Information Act (FOIA) requests to all 50 states and received responses from seven. Their publication includes summary tables describing the demographic composition of state DNA databases and annual DNA profile collection counts. In this work, we digitise and standardise those materials to enable integration with other public data to support structured reuse.

This Data Descriptor introduces three datasets that document the structure, scale, and composition of U.S. forensic DNA databases. The first is a national-level NDIS time series (2001--2025) reconstructed from archived FBI webpages, including monthly counts by index and jurisdiction, participating laboratory counts, and investigations aided. The second is a current cross-sectional SDIS dataset that compiles reported totals and profile-type breakdowns if applicable, with accompanying metadata on arrestee collection, familial search authorization, and relevant state statutes for all 50 states. The third dataset provides structured versions of two resources derived from Murphy and Tong (2020): 1) sex and racial composition by state, year, and index type, and 2) annual state-level profile collection counts.

These datasets are intended to support research on the scale and governance of forensic DNA infrastructure in the United States, including cross-state comparison, time series modeling, and evaluation of demographic composition. The longitudinal NDIS dataset also enables investigation of how operational or policy shifts---such as the expansion of the CODIS core STR set to 20 loci in 2017 \cite{Hares2015-ep}---may have influenced profile accumulation. All datasets are versioned, documented, and released with processing code to facilitate transparent reuse.

\section*{Methods}

\subsection*{Data Collection Strategy}

We developed a three-pronged approach to capture different dimensions of the U.S. forensic DNA database landscape. Federal statistics were reconstructed from archived FBI websites, state policies were compiled through targeted web searches of current legislative sources, and demographic data were obtained through Murphy \& Tong's supplementary data documenting FOIA responses on the demographic composition of seven states who responded to their requests.

\subsection*{Federal Statistics Extraction}

We queried the Internet Archive's Wayback Machine API to identify all preserved available archived snapshots of FBI NDIS statistics pages (Figure 1). Our search strategy accounted for the FBI's evolving web architecture: prior to 2007, statistics were distributed across jurisdiction-specific HTML pages (e.g., al.html for Alabama, ne.html for Nebraska), while post-2007 data were consolidated onto unified pages. We implemented a two-phase search algorithm that queried over 50 jurisdiction-specific URLs for the pre-2007 era and at least five distinct URL patterns for consolidated pages, testing both HTTP and HTTPS protocol variants to ensure comprehensive coverage across the archive's indexing variations.

Rather than sampling, we retrieved all available archived snapshots to maximize temporal resolution, yielding 11,359 unique captures spanning 255 distinct URL patterns. Each download attempt included robust error handling with exponential backoff (1s $\rightarrow$ 2s $\rightarrow$ 4s delays) and extended timeouts to manage network variability. This comprehensive approach captured the complete temporal evolution of NDIS reporting from July 2001 through August 2025 (Figure 2).

\begin{figure}[ht!]
\centering
\includegraphics[width=\textwidth]{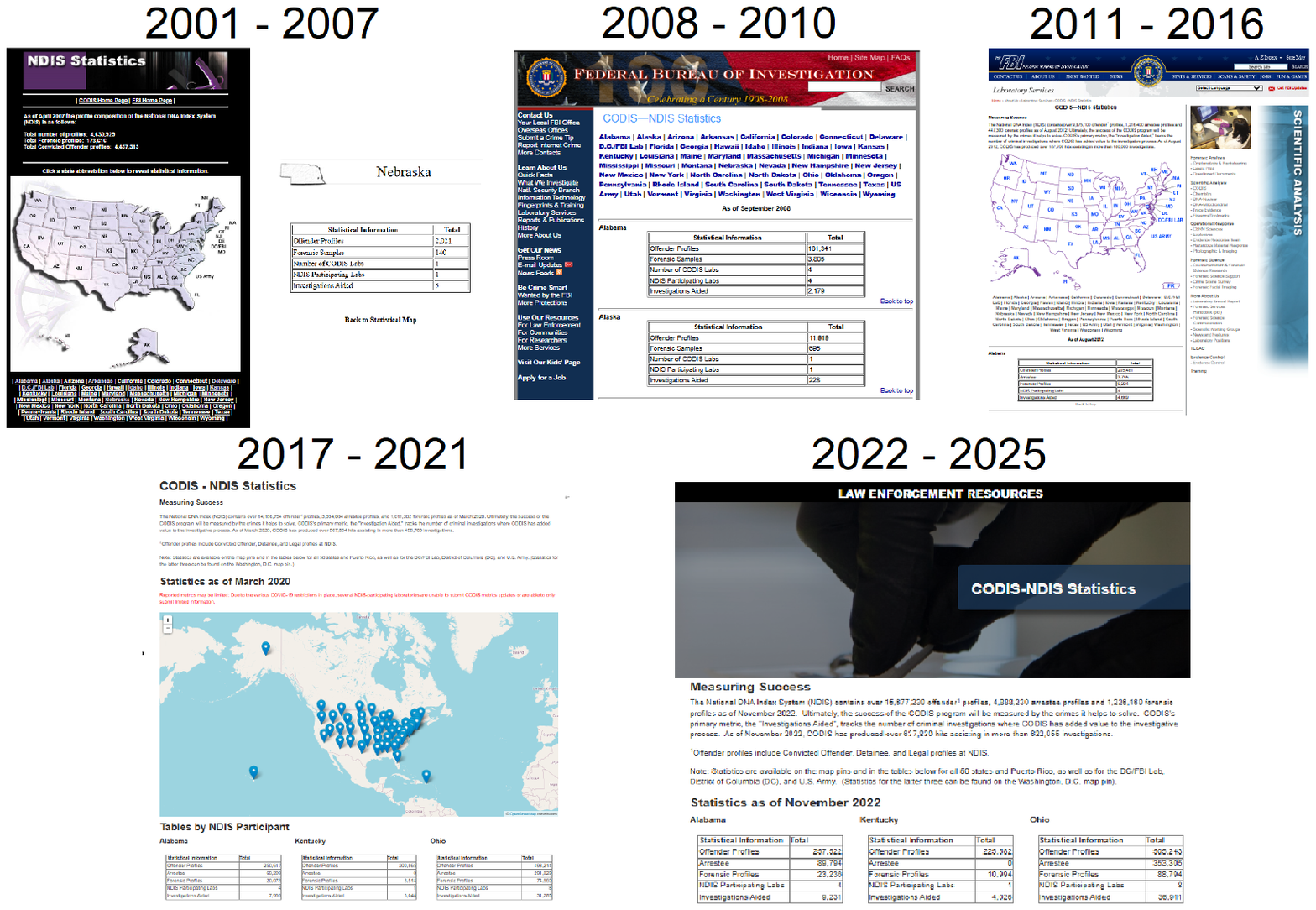}
\caption{\textbf{FBI formatting changes in NDIS statistics pages.} Variations in the structure and organization of FBI NDIS webpages captured through FBI.gov in different eras. Webpages are grouped into five eras used to build the NDIS time series dataset.  Eras of NDIS webpages are broken down into the pre-2007 era with jurisdiction-specific HTMLs,  unified HTMLs with jurisdiction-based dividers from 2008-2010, modified section headers in 2011-2016, and more modernized unified pages in the post-2017 eras.   }
\label{fig:ndis_eras}
\end{figure}

We processed downloaded available archived snapshots using four era-specific parsers designed to accommodate the FBI's evolving reporting formats. The pre-2007 parser extracted data from individual state pages, identifying jurisdictions through both filename patterns (e.g., extracting ``Pennsylvania'' from ``20040312\_pa.html'') and HTML content matching. Since these early pages lacked temporal metadata, we set report dates to null, relying instead on Wayback Machine capture timestamps. The 2008-2011 parser processed consolidated pages with jurisdiction sections separated by ``Back to top'' dividers and introduced extraction of ``as of'' date statements. The 2012-2016 parser captured the introduction of arrestee profile reporting alongside modified section headers. The post-2017 parser implemented regex patterns for the modern standardized format. Each parser extracted jurisdiction name, profile counts by category, laboratory counts, and investigations aided, along with temporal metadata from either explicit dates or snapshot timestamps.

The raw HTML data contained 169 unique text strings representing 54 actual jurisdictions, resulting from inconsistent naming conventions across time periods and data sources. We developed a comprehensive mapping table to standardize heterogeneous formatting of jurisdiction names while maintaining traceability to original sources. Examples included ``Alabama'', ``Alabama Alabama'', and ``Alabama Stats Alabama'' all representing the single jurisdiction of Alabama. Several non-state jurisdictions (i.e. Puerto Rico, DC/FBI Lab, DC/Metro PD, U.S. Army) were treated as distinct reporting entities and included when available.

\begin{figure}[ht!]
\centering
\includegraphics[width=\textwidth]{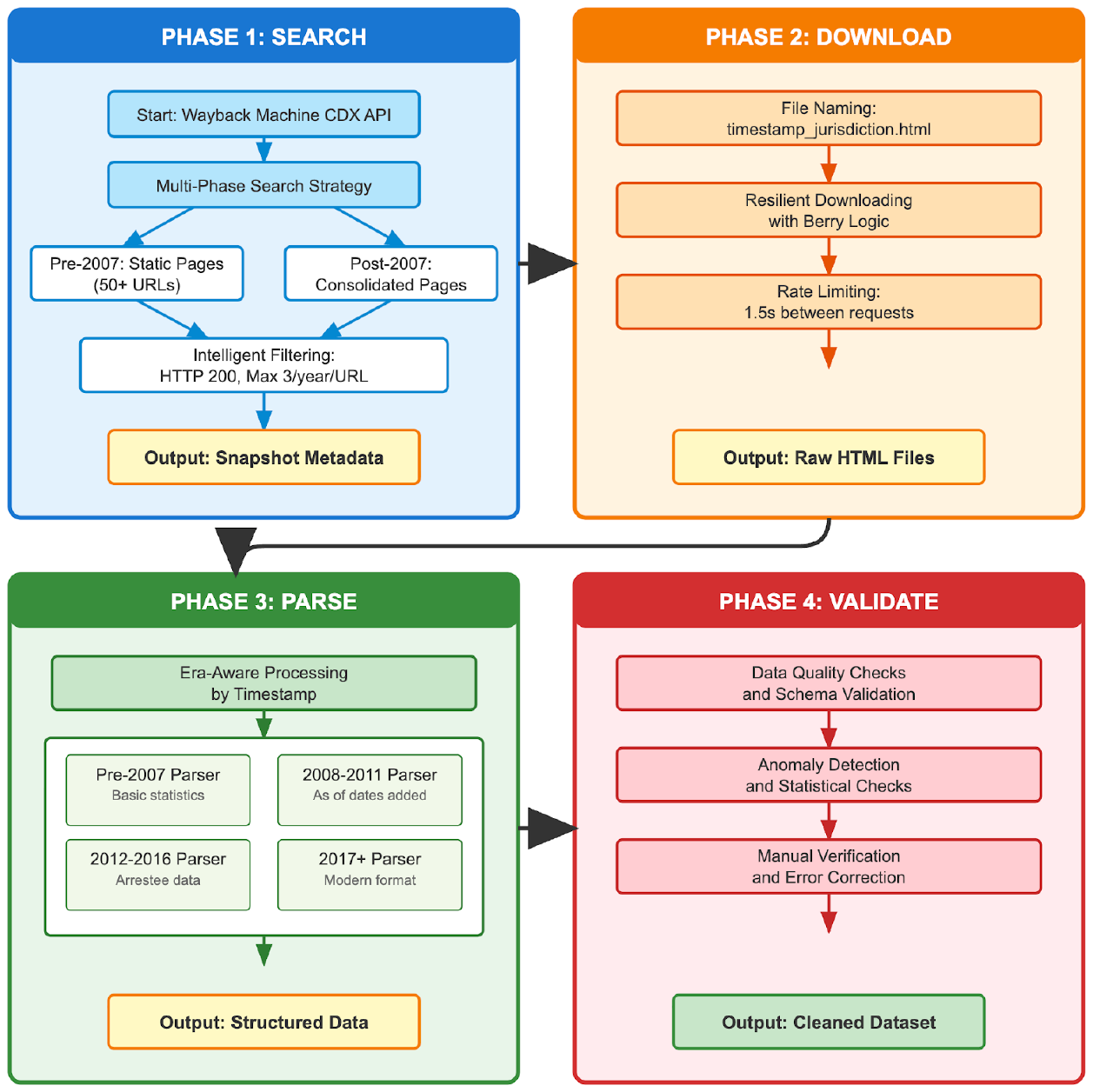}
\caption{\textbf{Parsing logic for NDIS statistics scraping.} A diagram depicting the step-by-step approach used to scrape the Wayback Machine and consolidate FBI NDIS statistics using era-specific parsers, allowing for a validated cleaned dataset of compiled NDIS time series data.}
\label{fig:parsing_logic}
\end{figure}

\subsection*{State DNA Database Counts and Policy Compilation}
Targeted web searches were performed to find available information on state-level DNA databases. Data sources were identified and reviewed using official state government websites and annual reports. Searches were conducted by entering terms such as ``SDIS,'' ``CODIS,'' ``Forensics'', or ``DNA'' into state website search bars, or by using general browser queries such as ``[state name] AND DNA Database,'' ``[state name] AND annual review,'' or ``[state name] Police AND DNA Database.'' Available information from state government webpages were codified to summarize the number of arrestees, offenders, forensic profiles, collection statutes and familial search practices per state. The most recent publicly available state-sourced database reports span from 2013 to 2025, depending on the state. Only 29 of the 50 states reported raw arrestee DNA totals. This represents the best available recent snapshot of state-level DNA database statistics. Policies regarding arrestee DNA collection were coded as binary variables (yes/no) when the statutory language was explicit (Figure 3A). Familial search policies were coded as "permitted" or "prohibited" when definitively stated or as ``unspecified'' when the statutes were silent or ambiguous on these particular provisions(Figure 3B).

\subsection*{Demographic FOIA Data Extraction and Standardisation}

We incorporated data from two appendices originally referenced in Murphy \& Tong (2020) that were not accessible directly with the original article. At our request, the authors provided the original files, which include the scanned FOIA response letters from state agencies reporting demographic breakdowns of their forensic DNA databases. Additionally, we provide an appendix with tables in which Murphy \& Tong had already performed calculations combining state-level collection data, Census demographics, and estimated rates of annual DNA collection by race.

To make these appendices accessible and reusable, we transcribed all documents into machine-readable CSVs. Analysis of FOIA responses revealed that state documents used "sex" and "gender" interchangeably. For data harmonization, we standardized on "sex" as the primary term, given that the presence of the Y sex chromosome is a key determinant of DNA profiling options. For the first appendix, this primarily involved converting the demographic tables from scanned PDFs into structured data while retaining the categories used by the original agencies. This data was available only for the seven states that responded to Murphy \& Tong's FOIA requests and the letters all date from 2018. For the second appendix, we treated Murphy \& Tong's own calculated estimates as data values, preserving them alongside the raw counts they reported. This data is also based on 2018. This ensured that our release captures not only the state-provided figures but also the derived metrics that underpinned the analysis in their original article.

Together, these appendices provide two complementary resources: 1) agency-reported demographic counts from seven states and 2) Murphy \& Tong's standardized estimates of annual DNA collection rates by race across all fifty states. We preserved the structure and terminology of the original materials while adding clear provenance fields so users can distinguish between reported values and author-calculated estimates. Users should note that the demographic data are limited to seven states and reflect 2018 reporting periods; these figures should not be treated as nationally representative or as indicative of current database composition.

\begin{figure}[ht!]
\centering
\includegraphics[width=\textwidth]{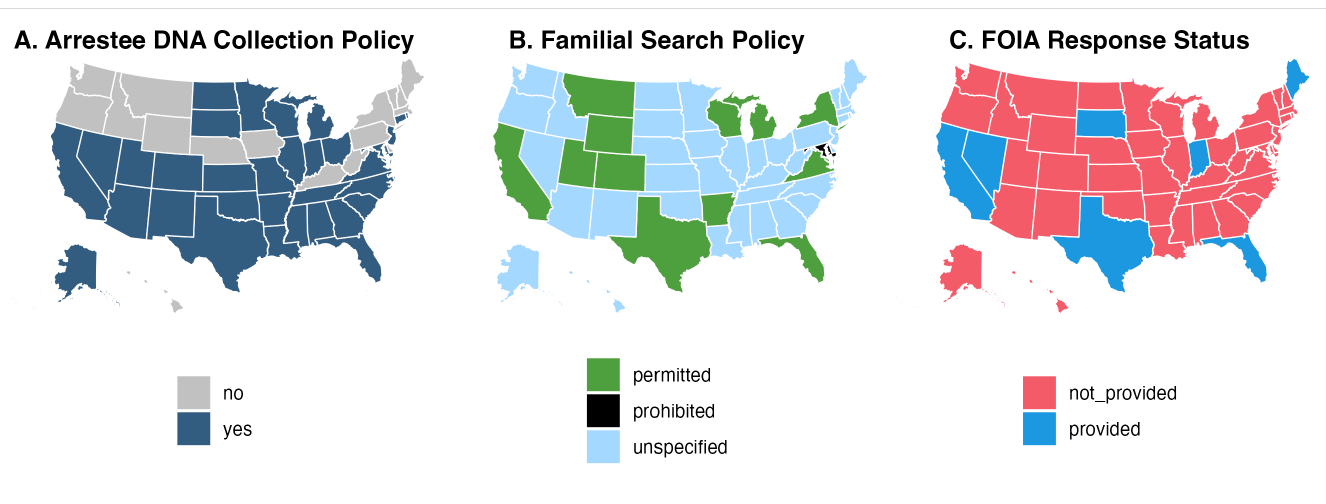}
\caption{\textbf{SDIS policies and FOIA availability mapped by state.} (A) Arrestee DNA Collection Policy availability coded as yes (blue) or no(gray) variables, (B) familial Search Policy detailing states with permitted(green), prohibited(black) and unspecified(blue) data availability, and (C) FOIA Response Status coding availability status as not\_provided(red) or provided(blue).}
\label{fig:sdis_map}
\end{figure}

\section*{Data Records}

The datasets presented in this study are available on Zenodo \cite{Pryor-Y-Donadio-J-P-Muller-M-Wilson-J-Lasisi-T2025-qq}. Detailed information and reproducible code used to process each of the datasets presented in this article, together with the full analyses and the scraping methodology, can be accessed through our accompanying research website: \url{https://lasisilab.github.io/PODFRIDGE-Databases/}. The raw NDIS time series, SDIS cross section, FOIA demographics and Annual DNA collection files are available for download in .csv format.

\begin{figure}[ht!]
\centering
\includegraphics[width=0.8\textwidth]{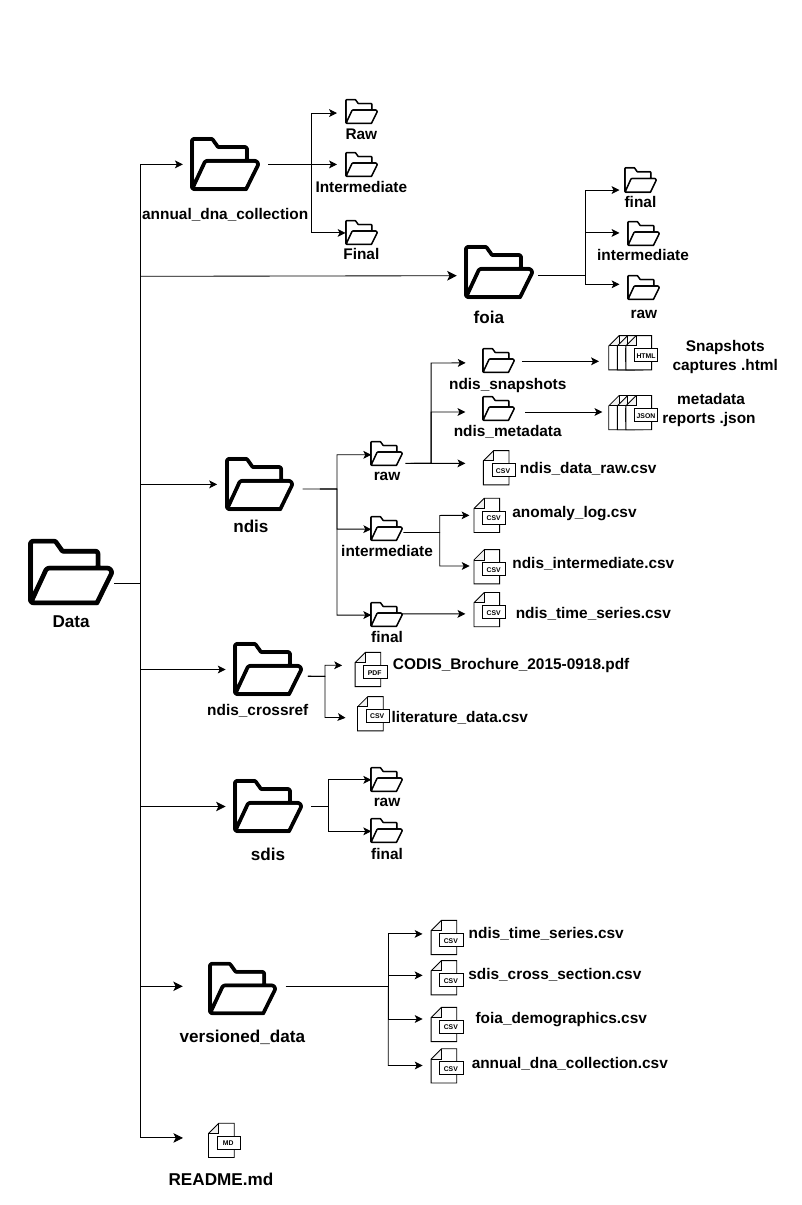}
\caption{\textbf{File and folder structure. }A file map showcasing the hierarchical organization of directories and data sources present in the public GitHub repository: \textbf{\url{https://github.com/lasisilab/PODFRIDGE-Databases}}}
\label{fig:file_structure}
\end{figure}

Figure \ref{fig:file_structure} presents a visual representation of the hierarchical directories that reflect the data sources, processing stages, and the final outputs derived. Each major data stream (NDIS, SDIS, FOIA, and Annual DNA Collection) is subdivided into standardized processing levels (Raw, Intermediate, and Final) to ensure transparency and reproducibility between analyses. The ndis\_crossref subfolder houses a 2015 CODIS brochure, as well as the data from Table \ref{tab:reference}. The definitions for all organizational terminology used across all files are available as a data dictionary in the README.md file located in the data subfolder and also located on our Github repository.

\section*{Technical Validation}

\subsection*{National DNA Index System (NDIS)}

We validated the NDIS time series by testing temporal consistency within each jurisdiction and at the national aggregate level. Because the counts of offender, arrestee, and forensic profiles, as well as investigations aided, are cumulative measures, they are expected to increase monotonically across successive data snapshots. Although occasional expungements may occur, such legitimate decreases should be distinguishable from reporting artifacts through systematic anomaly detection. Rather than removing anomalous values, we developed a flagging system that marks suspicious data points with boolean columns, preserving all original values and allowing users to apply their own filtering criteria. We encourage users to analyze both the flagged final data and the raw and intermediate files in our \texttt{data/ndis} folder (see Figure~\ref{fig:file_structure}), as they may wish to set their own definitions of what constitutes an anomaly.

To identify data-quality issues, we developed a sequential flagging pipeline. For each jurisdiction $j$, metric $x$, and capture time $t$, we compared the recorded count $N_{j,t}^{(x)}$ against the most recent \textit{unflagged} value (baseline) for that jurisdiction-metric pair. All spike and dip rules use a recovery window of at least 6 months, extended if needed to ensure at least 10 data points. This dual threshold ensures that transient errors (which should be corrected within months as the website is refreshed) are distinguished from genuine changes: if a value persists beyond the recovery window without returning to baseline, it is treated as a real change and the baseline is updated.

The four flagging rules are applied sequentially, so that values flagged by earlier rules are excluded from baseline comparisons in subsequent rules:

\textbf{1. Decimal spike-dip} (\texttt{flag\_decimal\_spike\_dip}). Applied to all metrics. This rule detects suspected decimal place entry errors by flagging values where $N_{j,t}^{(x)} > 8 \times \text{baseline}$ or $N_{j,t}^{(x)} < 0.125 \times \text{baseline}$, provided recovery toward the baseline occurs within the recovery window.

\textbf{2. Spike-dip} (\texttt{flag\_spike\_dip}). Applied to offender profiles, forensic profiles, arrestee profiles, and investigations aided. This rule flags sudden doublings or halvings: $N_{j,t}^{(x)} > 2 \times \text{baseline}$ or $N_{j,t}^{(x)} < 0.5 \times \text{baseline}$, again requiring recovery within the window.

\textbf{3. NDIS labs spike} (\texttt{flag\_ndis\_labs\_spike}). Applied to NDIS laboratory counts only. Because lab counts are small integers (typically 1--20), a lower threshold is used: $N_{j,t}^{(x)} > 3 \times \text{baseline}$ or $N_{j,t}^{(x)} < 0.33 \times \text{baseline}$.

\textbf{4. Stale exact reappearance} (\texttt{flag\_stale\_exact\_reappearance}). Applied to offender profiles, forensic profiles, and arrestee profiles. This rule flags instances where an exact earlier value reappears after higher values have been recorded, indicating cached or stale data served by the FBI website. Consecutive identical values are not flagged, as these represent legitimate unchanged counts between reporting periods.

All flags are preserved as boolean columns in the released dataset. This approach allows users to filter the data according to their own analysis needs---for example, excluding all flagged rows for a conservative analysis or retaining them to study reporting patterns.

\paragraph{Anomaly summary.}

\begin{figure}[ht!]
\centering
\includegraphics[width=\textwidth]{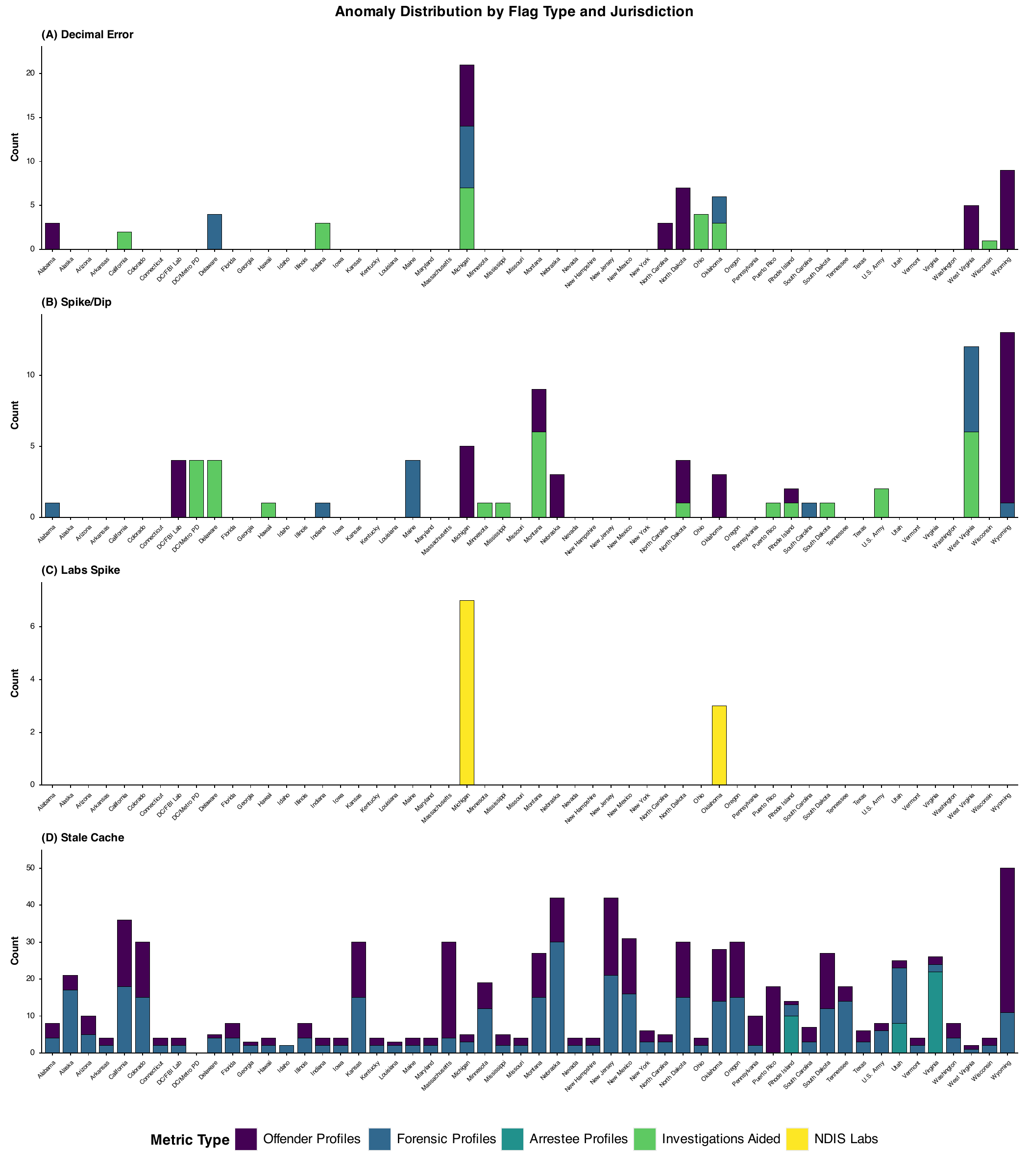}
\caption{\textbf{Anomaly flag distribution.} Four-panel stacked bar chart showing the distribution of data quality flags across jurisdictions. (A)~Decimal Error flags, (B)~Spike/Dip flags, (C)~Labs Spike flags, and (D)~Stale Cache flags. Bars are colored by metric type (Offender Profiles, Forensic Profiles, Arrestee Profiles, Investigations Aided, NDIS Labs). Each panel uses an independent $y$-axis to accommodate differences in flag counts across rule types.}
\label{fig:anomalies}
\end{figure}

\begin{figure}[ht!]
\centering
\includegraphics[width=\textwidth]{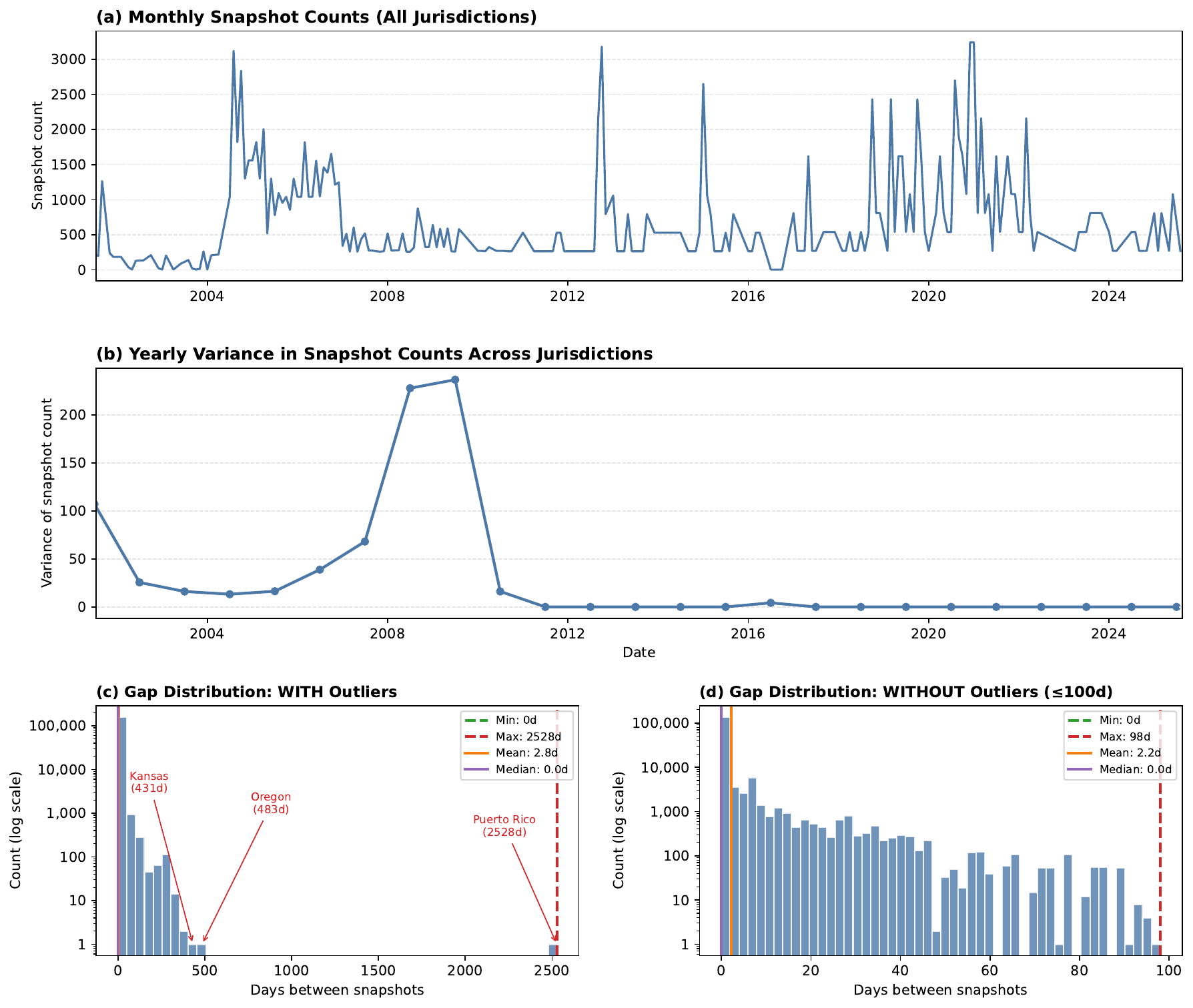}
\caption{\textbf{Archival coverage and time gap analysis.} (a)~Monthly snapshot counts across all jurisdictions, showing temporal density of archived captures. (b)~Yearly variance in snapshot counts across jurisdictions, illustrating how uniformly the Internet Archive captured NDIS pages over time. (c)~Distribution of time gaps between successive snapshots (all gaps), with annotated outliers including Puerto Rico (2,528~days), Oregon (483~days), and Kansas (431~days). (d)~Distribution of time gaps filtered to $\leq$100~days, showing the typical gap structure. Reference lines in (c) and (d) indicate minimum, maximum, mean, and median gap values.}
\label{fig:timegaps}
\end{figure}

Using these rules, we identified 898 flags across all metrics and jurisdictions (Figure~\ref{fig:anomalies}). The majority were stale exact reappearances ($743$), followed by spike-dip events ($77$), decimal spike-dip errors ($68$), and NDIS labs spikes ($10$). By metric, offender profiles had the most flags ($420$), followed by forensic profiles ($379$), investigations aided ($49$), arrestee profiles ($40$), and NDIS laboratories ($10$). Figure~\ref{fig:timegaps} shows the archival coverage of NDIS snapshots over time and the distribution of time gaps between successive captures, including annotated outlier jurisdictions with the largest gaps.

\paragraph{External calibration.}

\begin{figure}[ht!]
\centering
\includegraphics[width=\textwidth]{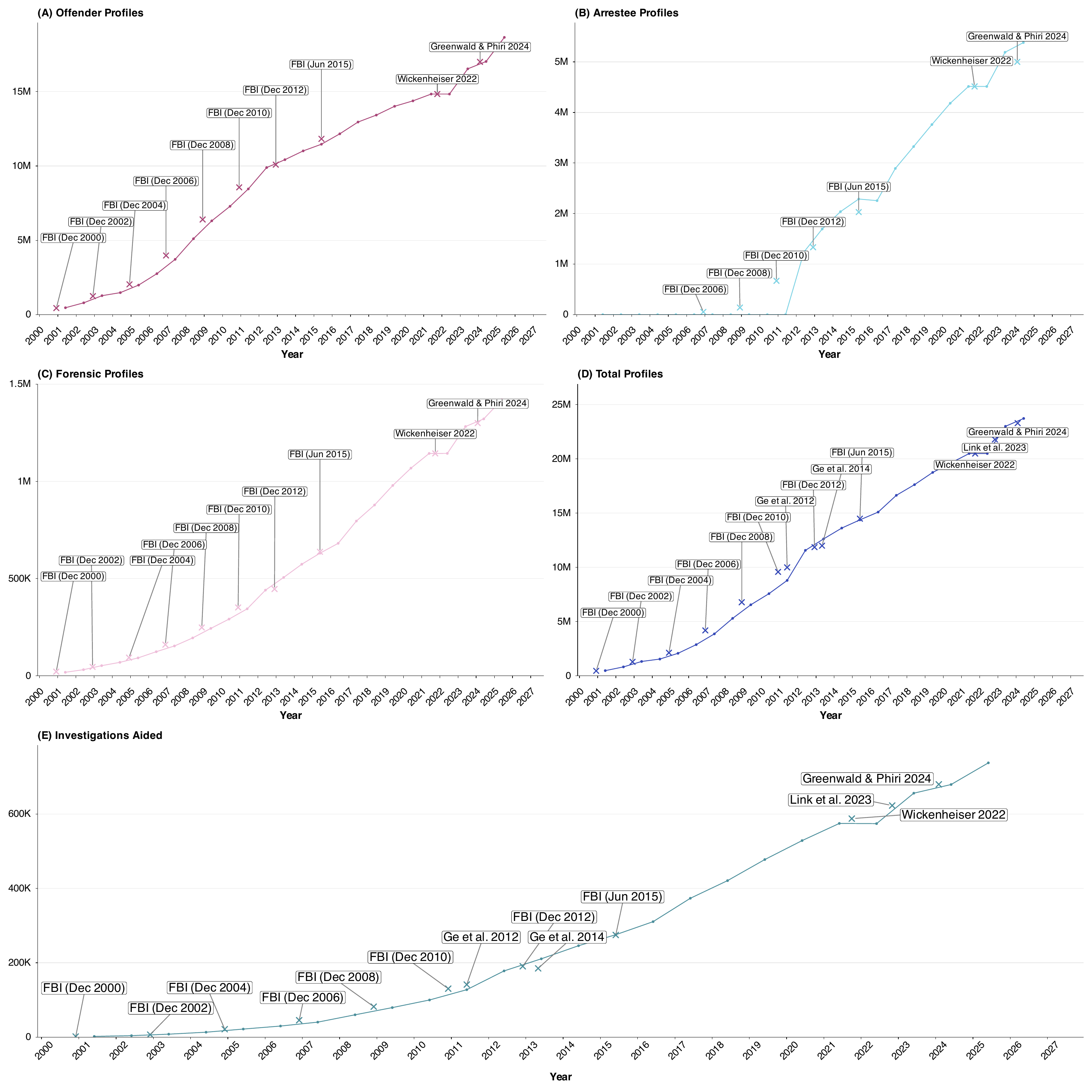}
\caption{\textbf{External calibration of NDIS data.}  NDIS data cross-referenced with both citations from the literature and numbers reported by an FBI 2015 CODIS brochure are plotted by year and split into five major metrics, (A)Offender Profiles, (B) Arrestee Profiles, (C) Forensic Profiles, (D) Total Profiles, and (E) Investigations Aided, showing NDIS time scaled growth from 2001-2025.}
\label{fig:external_calibration}
\end{figure}

After cleaning, we compared national totals to independent reference points from FBI CODIS brochures (2000–2015) and published estimates. For each year, national totals were computed by summing the maximum jurisdiction-level values for that year. The reconstructed offender, arrestee, forensic, total-profile, and investigations-aided series closely match values from a 2015 FBI Brochure (available in our Zenodo `ndis\_crossref` folder) and the point-in-time estimates reported by Ge et al.\ (2012, 2014), Wickenheiser (2022), Link et al.\ (2023), and Greenwald \& Phiri (2024) (Figure~\ref{fig:external_calibration}; Table~\ref{tab:reference}). This correspondence supports the fidelity of the reconstructed national trajectory.

\begin{table}[ht!]
\centering
\caption{\textbf{Reference values from FBI publications and academic literature for NDIS database growth.}}
\label{tab:reference}
\begin{tabular}{lcccccc}
\toprule
Citation & Date & Offender & Arrestee & Forensic & Total & Investigations \\
& & Profiles & Profiles & Profiles & Profiles & Aided \\
\midrule
FBI (Dec 2000) & 2000-Dec-01 & 441,181 & NA & 21,625 & NA & 1,573 \\
FBI (Dec 2002) & 2002-Dec-01 & 1,247,163 & NA & 46,177 & NA & 6,670 \\
FBI (Dec 2004) & 2004-Dec-01 & 2,038,514 & NA & 93,956 & NA & 21,266 \\
FBI (Dec 2006) & 2006-Dec-01 & 3,977,435 & 54,313 & 160,582 & NA & 45,364 \\
FBI (Dec 2008) & 2008-Dec-01 & 6,399,200 & 140,719 & 248,943 & NA & 81,955 \\
FBI (Dec 2010) & 2010-Dec-01 & 8,564,705 & 668,849 & 351,951 & NA & 130,317 \\
FBI (Dec 2012) & 2012-Dec-01 & 10,086,404 & 1,332,721 & 446,689 & NA & 190,560 \\
FBI (Jun 2015) & 2015-Jun-01 & 11,822,927 & 2,028,734 & 638,162 & NA & 274,648 \\
Ge et al. 2012 & 2011-Jun-01 & NA & NA & NA & 10,000,000 & 141,300 \\
Ge et al. 2014 & 2013-May-01 & NA & NA & NA & 12,000,000 & 185,000 \\
Wickenheiser 2022 & 2021-Oct-01 & 14,836,490 & 4,513,955 & 1,144,255 & NA & 587,773 \\
Link et al. 2023 & 2022-Nov-01 & NA & NA & NA & 21,791,620 & 622,955 \\
Greenwald \& Phiri 2024 & 2024-Feb-01 & 17,000,000 & 5,000,000 & 1,300,000 & NA & 680,000 \\
\bottomrule
\end{tabular}
\end{table}

\clearpage

\section*{Data Availability}

The datasets are archived on Zenodo: \url{https://doi.org/10.5281/zenodo.18820515}. Contents include NDIS (time series), SDIS (state cross-section and arrestee/familial search policies), and FOIA components. The FOIA folder contains the original scanned state responses received by Murphy \& Tong (2020) and machine-readable transcriptions. Reuse of the FOIA scans should cite Murphy \& Tong (2020) \cite{Murphy2020-pc} in addition to this dataset.

\section*{Code Availability}

All code used to generate the datasets is available in a public GitHub repository: \url{https://github.com/lasisilab/PODFRIDGE-Databases}. The repository contains Quarto notebooks, parsing scripts, and workflow documentation used to process the raw inputs into the released CSVs.

\section*{Acknowledgements}

We thank Erin Murphy for providing access to the original FOIA-acquired state-level demographic disclosures and appendices from her original publication.

\section*{Author Contributions}

\textbf{Conceptualization:} Lasisi, T. \\
\textbf{Data curation:} Donadio, J. P.; Lasisi, T.; Muller, S. C.; Ranka, V.; Wilson, J.; Pryor, Y. \\
\textbf{Methodology:} Lasisi, T.; Donadio, J. P. \\
\textbf{Validation:} Lasisi, T.; Donadio, J. P.; Ranka, V. \\
\textbf{Visualization:} Donadio, J. P.; Lasisi, T.; Ranka, V.; Pryor, Y. \\
\textbf{Writing -- original draft:} Lasisi, T. \\
\textbf{Writing -- review and editing:} Lasisi, T.; Donadio, J. P.; Muller, S.C.; Pryor, Y.; Ranka, V.; Wilson, J. \\
\textbf{Supervision:} Lasisi, T.

\section*{Competing Interests}

The authors declare no competing interests.

\section*{Funding}
This research did not receive external funding.

\bibliographystyle{unsrt}
\bibliography{references}

@ARTICLE{Ge2012-if,
  title     = "Developing criteria and data to determine best options for
               expanding the core {CODIS} loci",
  author    = "Ge, Jianye and Eisenberg, Arthur and Budowle, Bruce",
  journal   = "Investig. Genet.",
  publisher = "Springer Science and Business Media LLC",
  volume    =  3,
  number    =  1,
  pages     = "1",
  month     =  jan,
  year      =  2012,
  doi       = "10.1186/2041-2223-3-1",
  language  = "en"
}

@ARTICLE{Ge2012-ve,
  title     = "{DNA} based familial searching and related statistical issues",
  author    = "Ge, Jianye",
  journal   = "J. Forensic Res.",
  publisher = "OMICS Publishing Group",
  volume    =  3,
  number    =  8,
  pages     = "e112",
  year      =  2012,
  doi       = "10.4172/2157-7145.1000e112"
}

@ARTICLE{Link2023-dm,
  title     = "Microsatellites used in forensics are in regions enriched for
               trait-associated variants",
  author    = "Link, Vivian and Zavaleta, Yu{\'o}mi Jhony A and Reyes,
               Rochelle-Jan and Ding, Linda and Wang, Judy and Rohlfs, Rori V
               and Edge, Michael D",
  journal   = "iScience",
  publisher = "Elsevier BV",
  volume    =  26,
  number    =  10,
  pages     = "107992",
  month     =  oct,
  year      =  2023,
  doi       = "10.1016/j.isci.2023.107992",
  copyright = "http://creativecommons.org/licenses/by-nc-nd/4.0/",
  language  = "en"
}

@ARTICLE{Greenwald2024-sg,
  title    = "On Accountability: Genetic Tools for Justice and Injustice in
              Criminal Proceedings",
  author   = "Greenwald, Emily and Phiri, Linda",
  journal  = "J. Sci. Policy Gov.",
  volume   =  25,
  number   =  1,
  pages    = "0109",
  month    =  oct,
  year     =  2024,
  doi      = "10.38126/JSPG250109",
  language = "en"
}

@ARTICLE{Ge2014-rl,
  title     = "Future directions of forensic {DNA} databases",
  author    = "Ge, Jianye and Sun, Hongyu and Li, Haiyan and Liu, Chao and Yan,
               Jiangwei and Budowle, Bruce",
  journal   = "Croat. Med. J.",
  publisher = "Croatian Medical Journal",
  volume    =  55,
  number    =  2,
  pages     = "163--166",
  month     =  apr,
  year      =  2014,
  doi       = "10.3325/cmj.2014.55.163",
  language  = "en"
}

@MISC{Unknown2022-yi,
  title        = "{CODIS} Archive",
  booktitle    = "{FBI.gov}",
  month        =  jun,
  year         =  2022,
  howpublished = "\url{https://le.fbi.gov/science-and-lab/biometrics-and-fingerprints/codis-2}",
  note         = "Accessed: 2025-10-17",
  language     = "en"
}

@ARTICLE{Murphy2020-pc,
  title     = "The racial composition of forensic {DNA} databases",
  author    = "Murphy, Erin and Tong, Jun",
  journal   = "Calif. Law Rev.",
  volume    =  108,
  pages     = "1847--1930",
  year      =  2020
}

@MISC{Unknown2022-qe,
  title        = "{CODIS-NDIS} Statistics",
  booktitle    = "Law Enforcement {{FBI}.Gov}",
  month        =  jul,
  year         =  2022,
  howpublished = "\url{https://le.fbi.gov/science-and-lab/biometrics-and-fingerprints/codis/codis-ndis-statistics}",
  note         = "Accessed: 2025-9-26",
  language     = "en"
}

@ARTICLE{Wickenheiser2022-tb,
  title     = "Expanding {DNA} database effectiveness",
  author    = "Wickenheiser, Ray A",
  journal   = "Forensic Sci. Int. Synerg.",
  publisher = "Elsevier BV",
  volume    =  4,
  number    =  100226,
  pages     = "100226",
  month     =  apr,
  year      =  2022,
  doi       = "10.1016/j.fsisyn.2022.100226",
  copyright = "http://creativecommons.org/licenses/by/4.0/",
  language  = "en"
}

@MISC{UnknownUnknown-zg,
  title        = "Forensic {DNA} Education for Law Enforcement Decisionmakers",
  booktitle    = "National Institute of Justice",
  howpublished = "\url{https://nij.ojp.gov/nij-hosted-online-training-courses/forensic-dna-education-law-enforcement-decisionmakers/partial-matches/codis-hierarchy-cont}",
  note         = "Accessed: 2025-9-26",
  language     = "en"
}

@ARTICLE{Hares2015-ep,
  title     = "Selection and implementation of expanded {CODIS} core loci in
               the United States",
  author    = "Hares, Douglas R",
  journal   = "Forensic Sci. Int. Genet.",
  publisher = "Elsevier BV",
  volume    =  17,
  pages     = "33--34",
  month     =  jul,
  year      =  2015,
  doi       = "10.1016/j.fsigen.2015.03.006",
  language  = "en"
}

@misc{Pryor-Y-Donadio-J-P-Muller-M-Wilson-J-Lasisi-T2025-qq,
  title     = {United States forensic {DNA} databases: {NDIS}, {SDIS}, and {FOIA} datasets},
  author    = {Pryor, Yemko and Ranka, Virum and Donadio, Jo{\~a}o Pedro and Muller, Samantha C. and Wilson, Jenna and Lasisi, Tina},
  year      = {2026},
  publisher = {Zenodo},
  doi       = {10.5281/zenodo.18820515},
  url       = {https://doi.org/10.5281/zenodo.18820515},
  note      = {Data set, Zenodo. \url{https://doi.org/10.5281/zenodo.18820515}}
}

\end{document}